\documentclass[conference]{IEEEtran}

\usepackage{amsmath}
\usepackage{amssymb}
\usepackage{algorithm}
\usepackage{algorithmic}
\usepackage{enumitem}
\usepackage{url}
\usepackage[binary-units=true]{siunitx}
\usepackage{hyperref}
\usepackage{cleveref}
\usepackage{color}
\usepackage{booktabs}
\usepackage{verbatim}
\usepackage{multirow,bigdelim}
\usepackage[caption=false,font=footnotesize]{subfig}

\usepackage[colorinlistoftodos]{todonotes}

\usetikzlibrary{calc}

\newcommand{\groupG}{G}

\newcommand{\process}{\ensuremath{\operatorname{process}(m)}}
\newcommand{\self}{\ensuremath{\operatorname{self}}}

\graphicspath{{./images/}}

\begin{document}

\title{Arbitrary Length k-Anonymous\\ Dining-Cryptographers Communication}

\author{\IEEEauthorblockN{David Mödinger}
	\IEEEauthorblockA{\textit{Institute of Distributed Systems} \\
		\textit{Ulm University}, Germany \\
		david.moedinger@uni-ulm.de}
	\and
	\IEEEauthorblockN{Alexander Heß}
	\IEEEauthorblockA{\textit{Institute of Distributed Systems} \\
		\textit{Ulm University}, Germany \\
		alexander.hess@uni-ulm.de}
	\and
	\IEEEauthorblockN{Franz J. Hauck}
	\IEEEauthorblockA{\textit{Institute of Distributed Systems} \\
		\textit{Ulm University}, Germany \\
		franz.hauck@uni-ulm.de}
	}

\maketitle

\begin{abstract}
Dining-cryptographers networks (DCN) can achieve information-theoretical privacy.
Unfortunately, they are not well suited for peer-to-peer networks as they are used in blockchain applications to disseminate transactions and blocks among participants.
In previous but preliminary work, we proposed a three-phase approach with an initial phase based on a DCN with a group size of $k$ while later phases take care of the actual broadcast within a peer-to-peer network.
This paper describes our DCN protocol in detail and adds a performance evaluation powered by our proof-of-concept implementation.
Our contributions are 
(i) an extension of the DCN protocol by von Ahn for fair delivery of   arbitrarily long messages sent by potentially multiple senders,
(ii) a privacy and security analysis of this extension, 
(iii) various performance optimisation especially for best-case operation, and 
(iv) a performance evaluation.
The latter uses a latency of \SI{100}{\milli\second} and a bandwidth limit of \SI{50}{\mega\bit\per\second} between participants.
The interquartile range of the largest test of the highly secured version took $35\si{\second}\pm 1.25\si{\second}$ for a full run.
All tests of the optimized common-case mode show the dissemination of a message within $0.5\si{\second}\pm 0.1\si{\second}$.
These results compare favourably to previously established protocols for k-anonymous transmission of fixed size messages, outperforming the original protocol for messages as small as \SI{2}{\kibi\byte}.
\end{abstract}

\begin{IEEEkeywords}
	Network Protocol, Privacy Protocol, Dining Cryptographers, Peer-to-Peer Networking, Protocol Optimisation
\end{IEEEkeywords}

\section{Introduction}

Blockchains and their associated cryptocurrencies are popular in research and industry,
so popular that even people without technical proficiency invest considerable amounts of money in blockchain-based cryptocurrencies.
Most blockchains contain sensitive data of their users.
As blockchains are essentially public this data is easily accessible and readable by everyone.
Such data entails information on purchasing behaviour, credit balances, and how the money has been acquired~\cite{meiklejohn2013fistful,ron2013quantitative}.
To protect this sensitive information, new applications often provide unlinkable payments through the use of ring signatures~\cite{monerolab2014mysterious,kopp2017design} or zero-knowledge proofs~\cite{miers2013zerocoin,ben2014zerocash}.
Bitcoin and other existing blockchain systems received proposals for privacy-enhancing mechanisms, e.g.~\cite{ruffing2014coinshuffle,bonneau2014mixcoin}.
Unfortunately, these proposals mostly focus on privacy on the persisted blockchain, as every user can acquire and inspect this data.
This has been intensively investigated, e.g., for zcash~\cite{kappos2018anozcash}.

The dissemination of transactions within the peer-to-peer network connecting all participants received much less attention~\cite{koshy2014analysis,Biryukov2014deanobitcoin}.
Researchers investigating this aspect, recognised a lack of network privacy for blockchain systems, leading to the adoption of old and  the development of new protocols to provide network-layer protections.
For Bitcoin, Dandelion~\cite{Bojja2017dandelion} was proposed to tackle some of these issues.
Monero~\cite{monerolab2014mysterious} adopted Kovri, a variant of the onion-routing protocol called Invisible Internet Project, usually known as I2P~\cite{conrad2014survey}.

We proposed a design for a privacy-preserving protocol~\cite{moedinger2018anobroadcast} for stronger privacy guarantees.
The protocol requires groups of participants, which run a dining-cryptographers network to share a message in the local group, providing k-anonymity.
In a second and third phase, the protocol switches to a statistical privacy protocol to finally broadcast the message to the full network.

In ~\cite{moedinger2020trustcom}, we provided a realisation of this design called 3P3.
We applied a preliminary version of an arbitrary length k-anonymous dining-cryptographers network to realise the first of the three phases of 3P3.
We provided a preliminary evaluation of the security, privacy and expected performance, using a simulation of the protocol design.
This preliminary version lacks a rigorous analysis, an implementation and several optimisation for real-world applicability, as well as a comparison to similar solutions.

\subsection{Contributions}

In this paper we extend our preliminary proposal to a real-world k-anonymous transmission protocol of arbitrary length messages.
The contributions of this paper can be summarised as:

\begin{itemize}
\item We extend the k-anonymous message-transmission protocol by von Ahn et al.~\cite{vonahn2003kanonmessages} to arbitrarily long messages, including a privacy and security analysis.
\item We optimise the extended protocol for performance by reducing the overhead for the most common case.
\item We provide a proof-of-concept implementation of our new protocol.
\item We perform an extended evaluation of the protocol using our implementation, using a container setup with throughput-limited links.
\end{itemize}

The introduced protocol is not limited to the blockchain use-case and can be applied in various alternative use-cases.
The k-anonymous arbitrary length communication protocol can be used to replace fixed-length protocols~\cite{vonahn2003kanonmessages,wang2007} in many group-based privacy schemes.
Possible applications are journalist communication, e.g., for classified documents, supporting document-sized messages as well as comments or confidential messages between participating journalists, as envisioned by Dissent~\cite{Corrigan2010dissent}.

\subsection{Roadmap}

Starting with \Cref{sec:bg}, we discuss the necessary background for this paper by introducing notation and the application scenario.
\Cref{sec:dc} introduces the dining-cryptographers network and the protocol by von Ahn et al.~\cite{vonahn2003kanonmessages}.
\Cref{sec:relwrk} provides a short introduction to comparative related work for the later evaluation results.
In \Cref{sec:prot}, we describe our extension of the protocol to handle messages of arbitrary length.
The privacy and security properties of this new protocol are discussed in \Cref{sec:priv}.
\Cref{sec:opt} introduces our approach to improve the performance of the protocol.
Lastly, \Cref{sec:perf} provides an evaluation of the implementation before we conclude in \Cref{sec:conc}.

\section{Scenario and Notation}
\label{sec:bg}

In this paper, we propose a protocol for a group-wide broadcast, which can be used for implementing the first phase of a multi-phase privacy-preserving broadcast, e.g.~\cite{moedinger2018anobroadcast}, as it would be necessary for transaction dissemination in blockchain systems.
Our protocol requires groups of peers within an underlying peer-to-peer network.
We assume, for simplicity, that the network has no multiple edges or loops, and that the network remains connected, even if all malicious or faulty nodes are removed.
Otherwise, the network could be partitioned by malicious nodes and our protocol would be unable to reach all nodes.
Known properties of connectedness~\cite[Ch. 4]{jackson2010social} can consider a given rate of attackers for network creation. Therefore our assumption is warranted and not a strong restriction on the underlying network.

Channels between nodes should be authenticated, encrypted and integrity protected.
Such a channel can be realised using modern cryptographic schemes, e.g., via mutual transport layer security (MTLS).
Further, we require a public-private key pair for encryption with shared public keys among the group as well as an agreed-upon ordering within the group, e.g., ordered by public keys.

We further assume the network underlying our protocol takes care of basic network functions and management operations, e.g., re-establishing interrupted connections, authentication and encryption.
Especially group join and leave operations should be managed by alternative solutions, preferably protocols with enhanced privacy expectations~\cite{moedinger2020pixy}.

Lastly, the protocols make use of commitments.
A commitment \(C_r(x)\) over some data \(x\) is a unique, irreversible identifier, that can be validated to be created from \(x\) when \(r\) is known.
Commitments in this paper are homomorphic, i.e., given two commitments \(C_{r_1}(x_1)\) and \(C_{r_2}(x_2)\), it holds that \(C_{r_1}(x_1)\oplus{}C_{r_1}(x_1)=C_{r_1+r_2}(x_1+x_2).\)
Intuitively, if data is combined, the commitments on this data can be combined as well, to validate the combination.
We make use of the well known Pedersen commitments~\cite{pedersen1992} which fulfil this property efficiently.

Throughout this paper, we make use of standard notations when possible.
\Cref{tab:notation} gives an overview of less common but also used notations.

\begin{table}[htbp]
	\centering 
	\begin{tabular}{m{0.21\columnwidth}p{0.7\columnwidth}}
		Term & Meaning \\
		\toprule{}
		\(\groupG\) & All participants of a group.\\
		\(\process\) & Application logic to process the message \(m.\)\\
		\(C_r(x).\) & Commitment on \(x\), blinded by \(r.\)\\
		\(\{X\}_i\) & Value \(X\) encrypted under the public key of \(i.\)\\
		\(x =\sim \mathcal{U}(a,b)\) & Uniformly chose a random real \(x\in[a,b]\). \\
		\(i =\sim \mathcal{U}\{a,b\}\) & Uniformly chose a random integer \(a\leq i\leq b\). \\
		\(E = \sim \mathcal{U}_{n}\{S\}\) & Uniformly select \(n\) distinct random elements from \(S\).\\
		\bottomrule{}
	\end{tabular}
	\caption{Notations used in this paper.}
	\label{tab:notation}
\end{table}

\section{Basic Dining-Cryptographers Protocols}
\label{sec:dc}

The dining-cryptographers scheme provides an information-theoretically private way of sharing a message within a group.
In this section we discuss the constructions of dining-cryptographers protocols up to introducing the message-transmission protocol by von Ahn et al. that we use as a base for our own protocol.

\subsection{Dining Cryptographers Networks}

The basic protocol of the dining-cryptographers (DC) network by Chaum~\cite{chaum1988dining} allows for unconditional sender untraceability for sending a single bit within a group of $k$ participants.
This protocol has been extended in various ways to send longer messages.
One possibility is the sum-modulo-\(q\)-based implementation, given by \Cref{alg:DCN}.
In this more general form, the protocol allows a single participant to share a message in an unlinkable way.
The protocol requires secure channels to communicate and an upper limit for message lengths \(\ell\), all shorter messages will be padded to reach this length.
Note that the algorithm could also be written using the XOR operator, by replacing addition mod \(q\) with XOR.

Given a fixed-length message, e.g., by a modulus for modular arithmetic, all participants \(g_i\) decide on their message \(m_i\) to share.
If participants do not intend to send a message, they will prepare \(m_i=0\).
Only one participant is allowed to share a message \(m_i\not=0.\)
Each participant \(g_i\) splits its message \(m_i\) in $k$ random slices \(s_{i,j}\), so that \(m_i = \sum_{j=1}^{k} s_{i,j}\) and shares \(s_{i,j}\) with participant \(g_j\), whereas \(s_{i,i}\) remains with \(g_i\).
All participants combine every slice they receive and share the combination with all other participants.
Therefore, all participants will receive an aggregate of all slices produced by every other participant and can calculate the aggregation of all messages, which corresponds to \(\sum_i m_i\).

\begin{algorithm}[htbp]
\begin{algorithmic}
\renewcommand{\algorithmicrequire}{\textbf{Input:}}
\renewcommand{\algorithmicensure}{\textbf{Environment:}}
\REQUIRE Message \(m_{\self}\)
\ENSURE Participants \(g_i\) with \(i\in \{1\ldots k\}\), the executing node \(g_{\self}\), message length \(\ell\)

\STATE \(X = \begin{cases}
m_{\self} & \text{if } m_{\self} \text{ not empty}\\
0 & \text{else}
\end{cases}\)

\FOR{\(i\in \{1\ldots k\}\)}
\STATE \(s_{\self,i} = \begin{cases}
    \sim \mathcal{U}\{0,2^{\ell}-1\} & \text{if } i<k\\
    X - \sum_{j=1}^{k-1} s_{\self,j} & \text{if } i=k
\end{cases}\)
\ENDFOR

\STATE Send \(s_{\self,i}\) to \(g_i:i\not=\self\)

\STATE Wait until $g_{\self}$ received all \(s_{j,\self}\)
\STATE $S_{\self} = \sum_{j=1}^{k} s_{j,\self}$
\\\COMMENT{Note: This sum includes \(s_{\self,\self}\) which was not sent}
\STATE Broadcast \(S_{\self}\)

\STATE Wait until $g_{\self}$ received all \(S_{j}\)
\RETURN $\hat{X} = \sum_j S_{j}$
\end{algorithmic}
	\caption{Dining cryptographers protocol using addition modulo \(q\).}
	\label{alg:DCN}
\end{algorithm}

This simple implementation is information-theoretically secure against identifying the originator of a message, but it is not secure against interruption.
A participant can use a random message to create a collision with other messages, preventing other participants from reading the message, whereas the participant itself can actually retrieve the sent message.
Further, with a message complexity of \(\mathcal{O}(n^2)\) a dining-cryptographers network does not scale well to high numbers of particiants.

\subsection{k-Anonymous Message Transmission}

As noted before, DC networks do not scale well with the number of participants and can be prevented from making progress by malicious participants.
To create a practical application despite these weaknesses, von Ahn et al.~\cite{vonahn2003kanonmessages} realised a sender- and receiver-k-anonymous protocol for message transmission.
Note that this is no longer a broadcast protocol.
Instead it relies on assigning participants to a number of disjoint groups with only $k$ members each.
To achieve the sender-k-anonymity, the message is first shared anonymously within a source group \(\groupG_s\).
Afterwards, all participants of this group \(\groupG_s\) send the message to all participants of the target group \(\groupG_t\).

\begin{figure}[ht]
	\tikzstyle{gs1nodes}=[circle, draw, thin,scale=.7]
	\tikzstyle{gs2nodes}=[circle, draw, thin,scale=0.7]
	\tikzstyle{gtnodes}=[circle, draw, thin,scale=0.7]
	\tikzstyle{message}=[red, opacity=0.4]
	\tikzstyle{message_dir}=[->, red, opacity=0.6]
	\tikzstyle{legend}=[rectangle,scale=.7]
	\tikzstyle{legendbot}=[rectangle,scale=.5]
	
	\centering
	\resizebox{\columnwidth}{!}{%
		\begin{tikzpicture}[auto, thick]		
		
		\foreach \place/\x in {{(-2.5,0.5)/11},{(-2.5,1.5)/12},
			{(-1.5,1)/13}}
		\node[gs1nodes] (gs\x) at \place {};
		
		\foreach \place/\x in {{(-0.5,0.5)/21},{(-0.5,1.5)/22},
			{(0.5,1)/23}}
		\node[gs2nodes] (gs\x) at \place {};
		
		\foreach \place/\x in {{(2.5,0.5)/1},{(2.5,1.5)/2},
			{(1.5,1)/3}}
		\node[gtnodes] (gt\x) at \place {};
		
		\draw[thin] (gs11) edge (gs12);
		\draw[thin] (gs11) edge (gs13);
		\draw[thin] (gs12) edge (gs13);
		\draw[message,<->] (gs11) edge (gs12);
		\draw[message,<->] (gs11) edge (gs13);
		\draw[message,<->] (gs12) edge (gs13);
		
		\draw[thin] (gs21) edge (gs22);
		\draw[thin] (gs21) edge (gs23);
		\draw[thin] (gs22) edge (gs23);
		
		\draw[thin] (gt1) edge (gt2);
		\draw[thin] (gt1) edge (gt3);
		\draw[thin] (gt2) edge (gt3);
		
		\draw[message,->] (gs21) edge (gt1);
		\draw[message,->] (gs21) edge (gt2);
		\draw[message,->] (gs21) edge (gt3);
		\draw[message,->] (gs22) edge (gt1);
		\draw[message,->] (gs22) edge (gt2);
		\draw[message,->] (gs22) edge (gt3);
		\draw[message,->] (gs23) edge (gt1);
		\draw[message,->] (gs23) edge (gt2);
		\draw[message,->] (gs23) edge (gt3);
		
		
		
		\node[legend] at (-2.1,1){\textsc{$\groupG_s$}};		
		\node[legend] at (-0.1,1){\textsc{$\groupG_s$}};		
		\node[legend] at (2.2,1){\textsc{$\groupG_t$}};
		
		\node[legendbot] at (-2,0){\textsc{Within $\groupG_s$}};
		\node[legendbot] at (1,0){\textsc{$G_s$ to $G_t$}};
		\end{tikzpicture}}
	\caption{Visualisation of the two parts of the protocol by von Ahn et al., sharing within the group and the final transmission between the source group $G_s$ and target group $G_t.$}
	\label{fig:overview}
\end{figure}
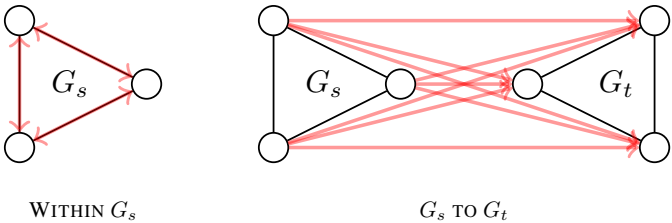

As the message does not need to reach all participants, the application to groups of sizes much smaller than all participants \(n\) comes naturally.
The protocol also builds upon other improvements~\cite{golle2004dcrev}, especially to identify cheaters through commitment schemes.
Thus, this new protocol deals with two severe weaknesses, even though it changed the use-case.

In the protocol by von Ahn et al. the dining-cryptographers phase shares an array of $2k$ slots, where each slot can consists of a message of fixed size and a non-zero group identifier, identifying the receiver group.
Each slot sums up to a total, fixed length of $\ell$.

In the commitment phase (see \Cref{alg:vACP}) participants prepare their message in a random slot and set all other slots to zero.
Further, they create commitments on all messages and broadcast these to the group.
The commitments allow for a blame protocol to identify malicious actors after detecting misbehaviour: if more than half of all slots are occupied, the one or multiple perpetrators can be uncovered.
If a sender detects a collision, i.e., their message is not correctly represented in the result, without less than half of all slots being occupied, it is deemed an accident instead of malicious action.
Note that all arithmetic is performed over \(\mathbb{Z}_q\) or, for commitments, the respective commitment structure.

\begin{algorithm}[H]
\begin{algorithmic}
\renewcommand{\algorithmicrequire}{\textbf{Input:}}
\renewcommand{\algorithmicensure}{\textbf{Environment:}}
\REQUIRE Message \(m_{\self}\), target group \(\groupG_t\)
\ENSURE Group \(\groupG_{\self}\) with \(|\groupG_{\self}|=k\), the executing node \(g_{\self}\), slot length \(\ell\)
\STATE Slot\( = \sim \mathcal{U}\{0,2k-1\}\)
\STATE \(X[i] = \begin{cases}
(m_{\self},\groupG_t) & \text{if } i=\text{Slot} \land \groupG_t \ne 0\\
(0,0) & \text{else }
\end{cases}\)

\FOR{\(t\in \{1\ldots 2k\}\) and \(i\in \{1\ldots k\}\)}
\STATE \(s_{\self,i}[t] = \begin{cases}
    \sim \mathcal{U}\{0,2^{\ell}-1\} & \text{if } i<k\\
    X[t] - \sum_{j=1}^{k-1} s_{\self,j} & \text{if } i=k
\end{cases}\)
\STATE \(r_{\self,i}[t] = \sim \mathcal{U}\{0,2^{\ell}-1\} \)
\STATE Compute \(\hat{\mathcal{C}}_{\self,i}[t] = C_{r_{\self,i}[t]}(s_{\self,i}[t])\)
\ENDFOR
\STATE Broadcast \(\{\hat{\mathcal{C}}_{\self,i}[t]: i\in \{1\ldots k\}, t\in \{1\ldots 2k\}\}\)
\end{algorithmic}
	\caption{Commitment phase of the k-anonymous message transmission protocol.}
	\label{alg:vACP}
\end{algorithm}

This results in a protocol with more overhead compared to basic dining-cryptographers networks but provides fairness and robustness in malicious environments.
It also allows for the transfer of multiple messages within one round.

\Cref{alg:vASP} is called the sharing round, where all computed information is shared.
\Cref{alg:vAB}, the local broadcast round, is the final group internal communication and allows all participants to reconstruct the messages with \Cref{alg:vAR}.
These \Cref{alg:vACP,alg:vAB,alg:vAR} are just the extension of the second half of \Cref{alg:DCN} by the validation and dissemination of commitments.
Finally, in \Cref{alg:vAT} each participant transmits the received message to every participant in the receiving group.

\begin{algorithm}[H]
\begin{algorithmic}
\renewcommand{\algorithmicrequire}{\textbf{Input:}}
\renewcommand{\algorithmicensure}{\textbf{Environment:}}
\REQUIRE $\forall i\in \{1\ldots k\}: s_{\self,i},r_{\self,i}$ of \Cref{alg:vACP}
\ENSURE Group \(\groupG\), the executing node \(g_{\self}\), commitments $\hat{\mathcal{C}}_{i,j}[t]$
\FOR{$g_i\in \groupG \setminus \{g_{\self}\}$}
\STATE Send \(\{(r_{\self,i}[t],s_{\self,i}[t]): t\in \{1\ldots 2k\}\}\) to \(g_i\)
\STATE Receiving node $g_i$ validates that $C_{r_{\self,i}[t]}(s_{\self,i}[t]) = \hat{\mathcal{C}}_{\self,i}[t]$
\ENDFOR
\end{algorithmic}
	\caption{Sharing round of the k-anonymous message transmission protocol.}
	\label{alg:vASP}
\end{algorithm}

\begin{algorithm}[H]
\begin{algorithmic}
\renewcommand{\algorithmicrequire}{\textbf{Input:}}
\renewcommand{\algorithmicensure}{\textbf{Environment:}}
\REQUIRE \(\forall j\in \{1\ldots k\}: (r_{j,\self}[t], s_{j,\self}[t])\) transmitted by others in \Cref{alg:vASP}
\ENSURE Group \(\groupG\), the executing node \(g_{\self}\), commitments $\hat{\mathcal{C}}_{i,j}[t]$
\STATE Wait until $g_{\self}$ received all pairs \((r_{j,\self}[t], s_{j,\self}[t])\)
\STATE $R_{\self}[t] = \sum_j r_{j,\self}[t]$
\STATE $S_{\self}[t] = \sum_j s_{j,\self}[t]$
\STATE Broadcast \(\{(R_{\self}[t],S_{\self}[t]):t\in \{1\ldots 2k\}\}\)
\STATE Everyone checks \(C_{R_{\self}[t]}(S_{\self}[t]) = \oplus_j \hat{\mathcal{C}}_{j,\self}[t]\)
\end{algorithmic}
	\caption{Broadcast within the group of the k-anonymous message transmission protocol.}
	\label{alg:vAB}
\end{algorithm}

\begin{algorithm}[H]
\begin{algorithmic}
\renewcommand{\algorithmicrequire}{\textbf{Input:}}
\renewcommand{\algorithmicensure}{\textbf{Environment:}}
\REQUIRE \(\forall j\in \{1\ldots k\}: (R_{j}[t], S_{j}[t])\) broadcasted by others in \Cref{alg:vAB}
\ENSURE Group \(\groupG\), the executing node \(g_{\self}\), commitments $\hat{\mathcal{C}}_{i,j}[t]$
\STATE Wait until $g_{\self}$ received all pairs \((R_{j}[t], s_{j}[t])\)
\STATE $R[t] = \sum_j R_{j}[t]$
\STATE $X[t] = \sum_j S_{j}[t]$
\STATE Everyone checks \(C_{R[t]}(X[t]) = \oplus_{i,j} \hat{\mathcal{C}}_{i,j}[t]\)
\RETURN $X$
\end{algorithmic}
	\caption{Result computation of the k-anonymous message transmission protocol.}
	\label{alg:vAR}
\end{algorithm}

\begin{algorithm}[H]
\begin{algorithmic}
\renewcommand{\algorithmicrequire}{\textbf{Input:}}
\renewcommand{\algorithmicensure}{\textbf{Environment:}}
\REQUIRE \(X,\) result from \Cref{alg:vAR}
\ENSURE The executing node \(g_{\self}\)
\FOR{\((m_i,\groupG_i) \in X\)}
\IF{\(G_i\not=0\)}
\STATE Send \(m_i\) to all members of \(\groupG_i\)
\ENDIF
\ENDFOR
\end{algorithmic}
	\caption{Transmission phase of the k-anonymous message transmission protocol.}
	\label{alg:vAT}
\end{algorithm}

The protocol by von Ahn et al. provides a notion of fairness: during honest execution, each participant has a probability of $p>\frac{1}{2}$ for successful transmission.
The protocol was described by von Ahn et al. with an extension for a zero-knowledge proof of fairness to ensure everyone used at most one slot.
This extension is not shown in the previous algorithms.
The proof is triggered when more than half of all slots are used and works in the following way:
A participant creates a permutation of all slots and a second set of commitments.
The verifier chose to either have the prover open all but one of the permuted commitments, i.e., showing that at most one commitment was not zero.
Alternatively, the verifier may choose to have the prover reveal the permutation and prove that they can open each corresponding commitments to the same value.

Lastly, von Ahn et al. provide an additional modification, which was not included in the previous description.
To handle selective non-participation, after preparing all message parts, all nodes must share any prepared message part with all participants.
The shared message parts are encrypted for the intended recipient.
If any participant claims another participant does not communicate with them, any participant can provide the encrypted backup of the message.
This vastly increases the overhead of the protocol, but is relevant for environments with highly expected malicious behaviour.

\section{Other Related Work}
\label{sec:relwrk}

In this section, we discuss privacy protocols for comparison to our proposal.
These especially include proposals with established performance evaluation for a comparison to our approach.

\subsection{Dissent}

Dissent~\cite{Corrigan2010dissent} is a protocol for sending a message anonymously in a small network, while the recipient may either be a single node or the whole group. 
Dissent provides high privacy guarantees, as it is based on a dining-cryptographers network.

Participants announce the length of their intended message to enable variable-sized messages.
The length announcements are kept anonymous via a secure random shuffle and encrypted values using onion encryption.
The participant removing the last layer of encryption publishes the shuffled lengths.
The length-dissemination phase~\cite{Corrigan2010dissent} scales linearly in the number of group members, e.g., a group size of 8 to 12 participants incurs a latency of around 30 seconds.

Dissent in Numbers~\cite{wolinsky2012dissentnumbers} is an extension of the original Dissent.
The protocol uses a small number of powerful core servers as anonymity providers in a round-based multi-phase protocol, to make it accessible for a large number of clients.

\subsection{k-Anonymous Overlay Protocol}

Wang et al.~\cite{wang2007} propose a peer-to-peer transmission protocol, implementing the same functionality as the k-anonymous message transmission protocol by von Ahn et al., i.e., point-to-point communication.

The network is partitioned in a number of rings, similar to Rac~\cite{mokhtar2013rac}.
Within each ring, nodes can communicate anonymously with each other.
This is achieved by a combination of layered encryption and shuffling batches of messages, similar to Dissents shuffle.
During each forwarding, every node must insert a message to prevent deanonymisation of participants.

If a node wants to send a message, they forward the ring identifier and message anonymously to another node in their own ring.
The receiving node forwards the message to a node in the target ring, unless the target is the local ring.
The receiving node broadcasts the message within the target ring, so that the recipient will receive the message.

The protocol claims to require a byzantine-secure setup phase to prevent malicious insertions in the ring configurations.
Further, the message transmission does not provide arbitrary-length messages, as their payload slots are fixed in size.

\section{Arbitrary Length Messages Protocol}
\label{sec:prot}

A major limitation of the k-Anonymous Message Transmission protocol proposed by von Ahn et al.~\cite{vonahn2003kanonmessages} is the pre-determined message size. If the selected message size is too small, it may prevent participants from disseminating messages that exceed that limit or if chosen to large, might create significant overhead if the message sizes vary singificantly in size. Our protocol addresses this limitation by allowing the participants to anonymously announce the size of their message in each round.  

\subsection{Overview}

Our protocol consists of two consecutive rounds, which are each built on a dining-cryptographers protocol.
The initial round determines whether there are senders that want to disseminate a message of a certain length. During the final round these messages are distributed.

As a first step during the initial round, participants with a message \(m\) first determines the length \(\ell\) of their message in Bytes. They further prepare a random round identifier \(r\) and a set $K$, composed of $k$ random values of fixed length.
The latter are required to ensure fairness during the actual message dissemination process in the final round, as we will point out later.
If a participant has no message to disseminate, they set all values to zero.
Finally, they share the set of values \((r,\ell,K)\) according to the protocol proposed by von Ahn et al.~\cite{vonahn2003kanonmessages}.
To achieve this, each participant with a tuple of non-zero values chooses a random slot in a pre-prepared vector with $2k$ slots, where they place their values. 
In case of four participants, this vector has the following form.

\begin{gather*}
[(r_1,\ell_1,K_1),
(r_2,\ell_2,K_2),
(r_3,\ell_3,K_3),
(r_4,\ell_4,K_4),\\
(r_5,\ell_5,K_5),
(r_6,\ell_6,K_6),
(r_7,\ell_7,K_7),
(r_8,\ell_8,K_8)]
\end{gather*}

The protocol then merges the vectors of all participants, using a secure multiparty computation to ensure that the contents of the individual vectors are hidden.
This is done by using the protocol of von Ahn et al.

During the final round, the actual messages are disseminated, similar to the approach implemented by Dissent~\cite{wolinsky2012dissentnumbers}.
Every participant prepares a second dining-cryptographers round for a message length equal to the sum of the message lengths announced during the initial round  \(\ell=\sum_{i=1}^{k} \ell_i\).
A participant \(g_i\) with a message length greater than zero determines their offset in the final message based on their random identifier $r_i$ and their announced length $\ell_i$.
Afterwards they fill their reserved section with their message \(m\).
The remaining portions of the message have to be set to zero.

%

The resulting message is then shared using a dining cryptographer variant.
The final inter-group transmission of von Ahn et al.'s protocol has not been adopted by our protocol since we only assume a single group at this point. 

\subsection{Initial Round}

The initial round of our protocol is used to share lengths of messages and prepare verifiable randomness for the final round.
The preparation protocol executed is given by \Cref{alg:r1}.
The group broadcast following this preparation is identical to \Cref{alg:vASP,alg:vAB,alg:vAR} of the protocol by von Ahn et al.~\cite{vonahn2003kanonmessages}.

For a fixed group size of \(k\) participants, we prepare \(2k\) slots for message lengths.
All slots consist of a random identifier \(r_i\), a message length \(\ell_i\) in Bytes and a set of \(k\) random values \(\{\{K_{i,j}\}_j : j \in \{1\ldots k\}\), encrypted to each participant \(g_j\).
When a participant \(g_i\) wants to disseminate a message \(m_i\) they chose a slot and a non-zero identifier \(r_i\) at random.

\begin{algorithm}[ht]
\begin{algorithmic}
\renewcommand{\algorithmicrequire}{\textbf{Input:}}
\renewcommand{\algorithmicensure}{\textbf{Environment:}}
\REQUIRE Message \(m\)
\ENSURE Group \(\groupG\) with \(|\groupG|=k\), the executing node \(g_{\self}\), public keys of group participants, slot length \(\ell\) in byte
\STATE \(r = \sim \mathcal{U}\{0,2^{16}-1\}\)
\STATE Slot\( = \sim \mathcal{U}\{0,2k-1\}\)
\STATE \(K = \{K_i = \{\sim \mathcal{U}\{0,2^{256}-1\}\}_{i} : i \in \{1\ldots k\}\}\)
\STATE \(X[i] = \begin{cases}
(r,\operatorname{length}(m), K) & \text{if } i=\text{Slot}\\
(0,0,0) & \text{else }
\end{cases}\)

\FOR{\(t\in \{1\ldots 2k\}\) and \(i\in \{1\ldots k\}\)}
\STATE \(s_{\self,i}[t] = \begin{cases}
    \sim \mathcal{U}\{0,2^{8\ell}-1\} & \text{if } i<k\\
    X[t] - \sum_{j=1}^{k-1} s_{\self,j} & \text{if } i=k
\end{cases}\)
\STATE \(r_{\self,i}[t] = \sim \mathcal{U}\{0,2^{8\ell}-1\} \)
\STATE Compute \(\hat{\mathcal{C}}_{\self,i}[t] = C_{r_{\self,i}[t]}(s_{\self,i}[t])\)
\ENDFOR
\STATE Broadcast \(\{\hat{\mathcal{C}}_{\self,i}[t]: i\in \{1\ldots k\}, t\in \{1\ldots 2k\}\}\)
\end{algorithmic}
	\caption{Preparation phase of the first DC round.}
	\label{alg:r1}
\end{algorithm}

The random identifiers $r_{\self}$ are required for a node to identify the slot of their message in the final round.
The length is not enough, as there might be multiple identical lengths.
After a successful execution of the initial round, each participant possesses a list of the form:

\begin{gather*}
\left[(r_1,\ell_1,\{K_{1,1}\}_1\ldots\{K_{k,1}\}_k\right),\\
\ldots,\\
\left(r_{2k},\ell_{2k},\{K_{1,2k}\}_1\ldots\{K_{k,2k}\}_k)\right].
\end{gather*}

\subsection{Final Round}

After the initial round has been successfully completed, i.e., the above list has been shared, all participants progress to the final round.
If the previous list contains at least one \(\ell_i\not=0\), everyone prepares a new compound message of length \(\sum_i \ell_i,\) the sum of all provided lengths.
Otherwise, the final round can be omitted and a new initial round is executed after a configurable time intervall.
The preparation is given by \Cref{alg:r2}, while the group broadcast is again identical to \Cref{alg:vASP,alg:vAB,alg:vAR}, but based on the output of \Cref{alg:r2}.

\begin{algorithm}[htbp]
\begin{algorithmic}
\renewcommand{\algorithmicrequire}{\textbf{Input:}}
\renewcommand{\algorithmicensure}{\textbf{Environment:}}
\REQUIRE Message \(m\), $r$ and $\ell$ from \Cref{alg:r1}, $X$ from previous round
\ENSURE Group \(\groupG\) with \(|\groupG|=k\), executing node \(g_{\self}\)

\FOR{\(t \in \{1\ldots 2k\}\)}
\STATE $(r[t],\ell[t],K[t]) = X[t]$
\IF{\(\ell[t]=0\)}
\STATE \textbf{continue}
\ENDIF
\STATE \(\hat{Y} = \begin{cases}
m & \text{if } r[t] = r \land \ell[t] =\ell\\
\underbrace{0\ldots{}0}_{\ell[t] \text{ many}} & \text{else }
\end{cases}\)
\FOR{\(i\in \{1\ldots k\}\) and \(\operatorname{part} \in \{1\ldots \lceil\frac{\ell[t]}{31}\rceil\}\)}
\STATE \(\hat{s}=\begin{cases}
    \sim \mathcal{U}\{0,2^{8\times 31}-1\} & \text{if } i<k\\
	\hat{Y}_{\operatorname{part}} - \sum_{j=1}^{k-1} s_{\self,j} & \text{if } i=k
\end{cases}\)
\\\COMMENT{Note: \(\hat{Y}_i\) references the \(i\)-th block of 31 Bytes of \(\hat{Y}\)}
\STATE \(\hat{r} = \operatorname{PRNG_{K[t][\self]}}(\sim \mathcal{U}\{0,2^{8\times 31}-1\}) \)

\STATE \(s_{\self,i}.\operatorname{append}(\hat{s})\)
\STATE \(r_{\self,i}.\operatorname{append}(\hat{r})\)
\STATE \(\hat{\mathcal{C}}_{\self,i}.\operatorname{append}\left(C_{\hat{r}}(\hat{s})\right)\)
\ENDFOR
\ENDFOR

\STATE Broadcast \(\{\hat{\mathcal{C}}_{\self,i}: i\in \{1\ldots k\}\}\)
\end{algorithmic}
	\caption{Preparation phase of the final round. Note that slots are variable in length and number, so they are only included, i.e., appended to a variable length list, when appropriate. For the same reason, commitments need to be generated for appropriate parts of correct length of the message, i.e., here 31 Bytes based on the base curve of the commitment scheme.}
	\label{alg:r2}
\end{algorithm}

Any node that submitted a length \(\ell\not=0\) has reserved \(\ell\) bytes in the next message, where the order is determined by the resulting layout of the initial round.
Assumed there is a non-zero length \(\ell_j\) in slot \(j\).
The start of this message inside of the single compound message transmitted in the final round can simply be computed by the length of all previous messages mentioned in slots \(i\) with \(i<j\), formally message \(j\) starts at \(1+\sum_{i=1}^{j-1} \ell_i\).
All other parts not used by the message are set to zero.

\begin{figure}[H]
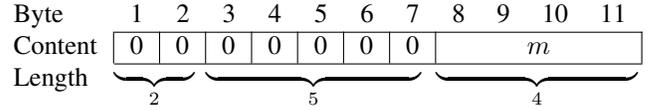

\centering
\begin{tabular}{|l|c|c|c|c|c|c|c|c|c|c|c|}
\multicolumn{1}{l}{Byte} &
\multicolumn{1}{c}{1} &
\multicolumn{1}{c}{2} &
\multicolumn{1}{c}{3} &
\multicolumn{1}{c}{4} &
\multicolumn{1}{c}{5} &
\multicolumn{1}{c}{6} &
\multicolumn{1}{c}{7} &
\multicolumn{1}{c}{8} &
\multicolumn{1}{c}{9} &
\multicolumn{1}{c}{10} &
\multicolumn{1}{c}{11} \\
\cline{2-12}
\multicolumn{1}{l|}{Content} &
0 & 0 & 0 & 0 & 0 & 0 & 0 &  \multicolumn{4}{c|}{$m$}\\
\cline{2-12}
\multicolumn{1}{l}{Length} &
\multicolumn{2}{@{}l@{}}{%
      \raisebox{.5\normalbaselineskip}{%
      \rlap{$\underbrace{\hphantom{\mbox{$1$\hspace*{\dimexpr4\arraycolsep+\arrayrulewidth}$2$}}}_{2}$}}%
    } &
\multicolumn{5}{@{}l@{}}{%
      \raisebox{.5\normalbaselineskip}{%
      \rlap{$\underbrace{\hphantom{\mbox{$1$\hspace*{\dimexpr4\arraycolsep+\arrayrulewidth}$2$\hspace*{\dimexpr4\arraycolsep+\arrayrulewidth}$3$\hspace*{\dimexpr4\arraycolsep+\arrayrulewidth}$4$}}}_{5}$}}%
    } &
\multicolumn{4}{@{}l@{}}{%
      \raisebox{.5\normalbaselineskip}{%
      \rlap{$\underbrace{\hphantom{\mbox{$1$\hspace*{\dimexpr4\arraycolsep+\arrayrulewidth}$2$\hspace*{\dimexpr4\arraycolsep+\arrayrulewidth}$3$\hspace*{\dimexpr4\arraycolsep+\arrayrulewidth}}}}_{4}$}}%
    } \\
\end{tabular}
\caption{Example message allocation in the final round. The first row gives the index of the bytes, the second row the occupation using the message lengths as an indicator and the last row shows the actual usage of the bytes.}
\label{fig:example}
\end{figure}

Assume an example network with 4 participants.
Then there will be 8 slots in the initial round.
Let us further assume that the transmitted lengths are 2, 5 and 4, in the order of appearance in these slots, e.g. \((2,0,0,5,4,0,0,0).\)
The participant having sent length 4, will then put its message from the eleventh to the fourteenth byte, as seen in \Cref{fig:example}.
All other bytes are set to zero.

For every message slot \(j\) not designated for node \(i\), they need to decrypt the seed value \(\{K_{i,j}\}_i\) addressed to them.
This seed must be used to deterministically create the randomness for all commitments for this slot \(j\).
The creator of this slot can later validate that the commitments are commitments to zero.
As all values should be zero, this has no implications for the privacy of the node creating the commitments.

After a successful transmission, all participants receive the same \([m_1,\ldots,m_i].\)
This can easily be verified.
For the slots, it holds that:

\begin{align*}
  &Y_{result}
    \stackrel{Alg.~\ref{alg:vAR}}{=} \sum_j S_j \stackrel{Alg.~\ref{alg:vAB}}{=} \sum_j \sum_h s_{h,j} \\
  	&\stackrel{}{=} \sum_h \sum_j s_{h,j} \stackrel{(*)}{=} \sum_h \left(s_{h,k} + \sum_{j\in \{1\ldots k-1\}} s_{k,j}\right) \\
  	& \stackrel{Alg.~\ref{alg:r2}}{=} \sum_h \left((\hat{Y}[h] - \sum_{j\in \{1\ldots k-1\}} s_{k,j}) + \sum_{j\in \{1\ldots k-1\}} s_{k,j} \right)\\
  	&= \sum_h \hat{Y}[h] = \sum_h [0,\ldots,0,m_h,0,\ldots,0] \\
  	&= [m_1,\ldots,m_i].
\end{align*}

Where \((*)\) just removed one entry from the sum.

Once the sender of a message notices a collision, i.e., their own message is damaged, they verify that all commitments of participants are correct.
If a commitment of participant \(i\) for slot \(j\) does not correctly validate to zero, the sender creates a blame message \((i,K_i,j,\text{round-offset}).\)
Here, the round offset identifies the instance of the protocol where the violation occurred, e.g., \(1\) for the previous instance.
This message is inserted in the next round, allowing other participants to validate the blame without revealing the sender of the message.

\section{Privacy and Security}
\label{sec:priv}

In this section, we discuss the privacy and security of our protocol.
Here, privacy concerns itself with how to identify the originator of a given message.
Security, on the other hand, covers the correctness and robustness of the protocol, i.e., if the protocol can be prevented from sharing the message.

\subsection{Initial Round}

Since the initial round is based on the k-anonymous message-transmission protocol proposed by von Ahn et al.~\cite{vonahn2003kanonmessages}, it exhibits the same correctness, robustness, fairness, and anonymity and is secure in the discrete logarithm model.
Here, robustness means that either all honest participants that have a message to disseminate will eventually successfully transmit their message, or an attacker is exposed.
To prevent a denial-of-service attack in the final round, the message lengths should be restricted, e.g., $2^{16}$ Bytes, when used in blockchain applications. Otherwise, an attacker could announce arbitrary message lengths, which would drastically slow-down the execution of the final round.

\subsection{Final Round: Security}

During the final round, the random blinding factors \(r\) for the commitments are generated using a pseudo random number generator seeded with \(K_i\) distributed during the initial round.
Therefore, the legitimate message sender can validate the commitments as zero commitments in case a collision occurs, since they can deterministically recompute all blinding factors \(r\) based on the \(K_i\) they sent, allowing them to check the commitment to zero.
If a commitment does not reveal to be zero, the legitimate message sender will inject a blame message \((\operatorname{\#round},i,K_i)\) in a later protocol instance.

Other participants can verify the seed and validate that the commitment is not zero.
If the accusation is true, they exclude the attacker.
An honest participant can not be blamed without breaking the security assumption of the underlying commitment scheme.
Therefore, either all messages are successfully transmitted or an attacker is identified and excluded.

The protocol is capable of dealing with non-cooperation, similar to the approach proposed by von Ahn et al.~\cite{vonahn2003kanonmessages}, by sharing encrypted instances of all \(\{(s_{\self,j},r_{self,j})\}_j\) pairs with every participant, allowing the reconstruction of the contents with selective non-cooperation.
This creates considerable additional load, so forming of a completely new group might be more economical.

Given the previous conclusions, the final round provides robustness properties similar to those of the initial round: either the transmission succeeds, or an attacker is excluded.
Therefore, the protocol will eventually succeed, for a finite number of attackers and sufficiently many honest participants.

\subsection{Final Round: Privacy}

By construction, i.e., using the same dissemination mechanism as the initial round, the protocol fullfils the privacy requirements as other dining-cryptographers protocols.
Nonetheless, we would like to highlight the privacy properties of important situations and how they come to be.

Revelations over recent years have shown that service providers and intelligence agencies across the world collect and analyse information, coming reasonably close to a global passive attacker.
If such an observer could collect all slices \(s_{i,j}\) of a participant, they could recompute the original message sent by the participant.
This is prevented by authenticated encrypted channels between participants.

In order to prevent traffic correlation, DC networks require all participants to send the same amount of data during each round, to ensure that the senders cannot be identified based on the amount of generated network traffic. While a global passive attacker can detect the communicating DC groups and the broadcast message, they can not identify the originator within the group~\cite{vonahn2003kanonmessages,chaum1988dining}.
Our protocol provides \(k\)-anonymity against this type of attacker, with \(k=|\text{group}|\).

Therefore, to improve the detection of DC group participants, an attacker has to be part of the group.
For \(\beta\) attackers within the group, dining cryptographer network-based communication trivially provides \((k-\beta)\)-anonymity: removing the \(\beta\) known keys \(s\) of attackers from the key-graph results in a remaining graph of size \(k-\beta\), cf. Chaum~\cite{chaum1988dining}.

The anonymity guarantees, i.e., the expected number of attackers \(\beta\), depend on the group formation mechanism~\cite{vonahn2003kanonmessages}.
Current strategies of a random selection of participants with an assumed attacker probability \(p\) require group sizes of \(\frac{2k}{1-p}\) to provide \(k\)-anonymity with high probability~\cite{vonahn2003kanonmessages}.
These bounds depend on the attacker probability distribution, which could be improved through external trust information during the group formation. 

\section{Performance Optimisation}
\label{sec:opt}

In this section, we discuss a multitude of optimisations built into the protocol, to improve its performance.

\subsection{Small Optimisations}

The protocol allows for various smaller optimisations to improve performance.

\textbf{Precomputation:}
To reduce startup latency, commitments can be prepared before a protocol instance is started or during the downtime of a previous instance.
The random blinding factors for the commitments of the initial round can be pre-generated.
Since each participant has to generate at least $2k-1$ zero commitments for the empty slots, these commitments can be precomputed as well.

\textbf{Deferred validation:}
While the validation of commitments is important to identify misbehaving nodes, it is sufficient to validate the commitments only when malicous behaviour is observed.
To achieve this, messages and commitments should be stored on disk for later validation.
Additionally, hash-based message codes augment the sent elements, i.e., the \((r,\ell,K)\) tuples in the initial round and messages in the final round, to detect misbehaviour more reliably.
A sender can always identify previously unnoticed misbehaviour and trigger validation.

\textbf{Direct transmission:}
If the message \(m\) a participant wants to send is short, i.e., shorter than the elements of the length message \(|m|<|(r,\ell,K)|,\) the message can replace the length identifier and be sent directly.
To allow this behaviour, as well as blame messages, some special values for \(r\) can be reserved to indicate a direct transmission.

\textbf{Size of the Random Identifier:}
To keep the size of random identifiers small, we do a quick evaluation based on the birthday paradox formula.
It provides a rough boundary, given a collision probability \(p\) and group size \(k\):

\[\operatorname{sizeof}(r)\geq \log_2\left(\frac{1}{1-(1-p)^{\frac{2}{k(k-1)}}}\right).\]

For a collision probability of \(1\%,\) we can tolerate 36 participants using 16 bits per random identifier.
As collisions should already be rare due to requiring the same length.
Most applications should function well with these values.

\subsection{Overhead Reduction in Normal Operation}

The proposed protocol has massive overhead to secure its operation in a malicious environment, e.g., commitments, seeds for commitments and more.
This overhead can be severely reduced in an environment where maliciousness is the exception, i.e., in normal operation, there is likely no attack on the robustness of the system.
If a possible attack is detected, e.g., multiple collisions or garbage results in a round, the protocol can still switch to a secure mode.
In its secure mode, the protocol is used as described in \Cref{sec:prot}.

The unsecured variant removes several aspects.
No commitments are created or added to the protocol.
Therefore, no commitment seeds need to be generated and transmitted in the initial round.
This results in a message of the form \((r_1,\ell_1),\ldots,(r_{2k},\ell_{2k})\) which can be encoded as \(2k\) integers of 4 or 8 bytes.
It's important to note that this removal of commitments does not impact the privacy of the dining cryptographer construction.

\begin{figure}[ht]
\includegraphics[width=\columnwidth]{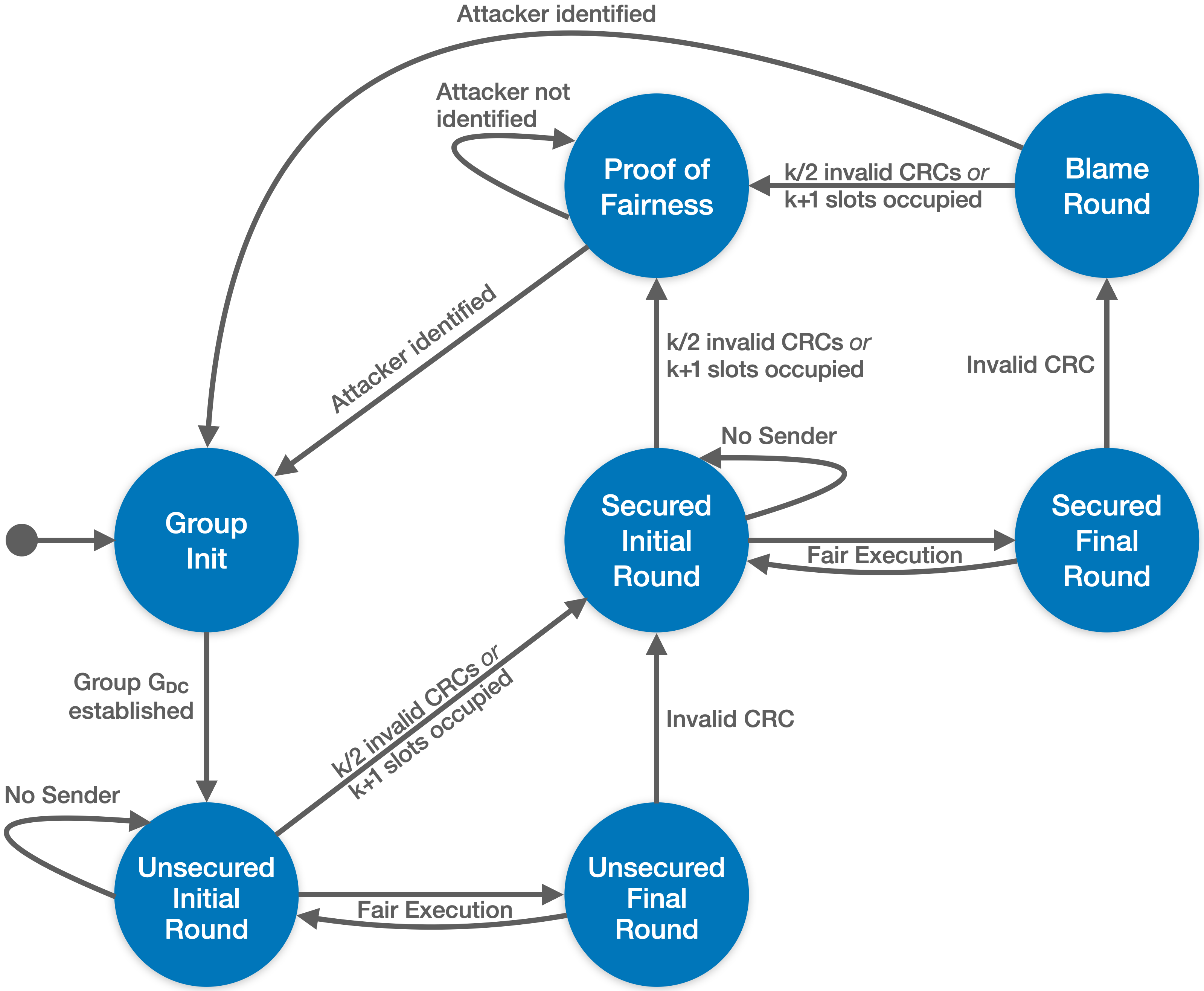}
\centering
\caption{State machine of the optimized protocol.}
\label{fig:statemachine}
\end{figure}

This construction results in a more complex protocol state machine, which is shown in \Cref{fig:statemachine}.
The group init phase represents the (re)formation of the group, possibly expelling identified attackers.
Once a group key \(G_{PC}\) is established, the protocol starts with unsecured rounds.
Once a likely attack is detected, i.e., more slots are occupied than there are participants or there are many collisions, the protocol will switch to a secure variation.

If the attacker correctly participates in the secured round, this will impose the aforementioned overhead without any attacker being revealed.
Depending on the scenario, instead of permanently operating in the secured mode, it may be advisable to create a group with fresh participants.

\section{Performance}
\label{sec:perf}

In this section, we discuss the performance evaluation of the protocol via the proof-of-concept implementation.

\subsection{Methodology}

In order to evaluate the performance of our protocol, we build a prototype implementation in C++.
For the performance evaluation, multiple instances of the implementation were deployed in separate docker containers on a single machine with 32 physical cores and 2 threads per core.
Each experiment involved 100 iterations of the protocol execution, to improve confidence in the results.
The source code and the experiment scripts are available as open-source software on GitHub\footnote{\url{https://github.com/vs-uulm/3p3-evaluation}}, including an explanation on how to reproduce the results.
We used the optimised secure variant with prepared commitments and omission of commitment validation during successful rounds for the evaluation of the secure version.

We introduced an artificial network latency of \SI{100}{\milli\second} and used traffic control to limit the bandwidth of each interface to 50Mbit/s in order to achieve more realistic results.
Message sizes were chosen to be fixed at \SI{512}{\byte} as an approximation of average \textit{Bitcoin} transaction sizes.

An additional docker container served as a central instance, which managed the connection setup and logging of all other deployed instances.
We performed a real-time benchmark of the relevant cryptographic operations utilised by the protocol on the machine, and observed runtimes of \SI{0.0109}{\milli\second} for point addition and \SI{0.6698}{\milli\second} for scalar multiplication. Combined, this results in a runtime of \SI{1.35}{\milli\second} per commitment generated over the secp256k1 curve, which involves two scalar multiplications and a single point addition operation.

\subsection{Number of Participants}

In order to compare the performance of the unsecured and secured protocol versions, we looked at the scaling behaviour based on the number of participating nodes.
In order to circumvent collisions during in the initial round, each sending node chose their slot based on their node ID. This modification has been applied for the evaluation purposes only, and should not be utilised in production systems, since it spoils the privacy guarantees of the protocol.

We scaled the group sizes from 8 to 24 participants in increments of 2 for both the secured and unsecured version.
We fixed the number of available threads for each instance to four and also used a fixed number of senders also set to four.
This experiment is codified in \texttt{nodes.sh} in our code repository.

\begin{figure}[htbp]
\centering
\includegraphics[width=\columnwidth]{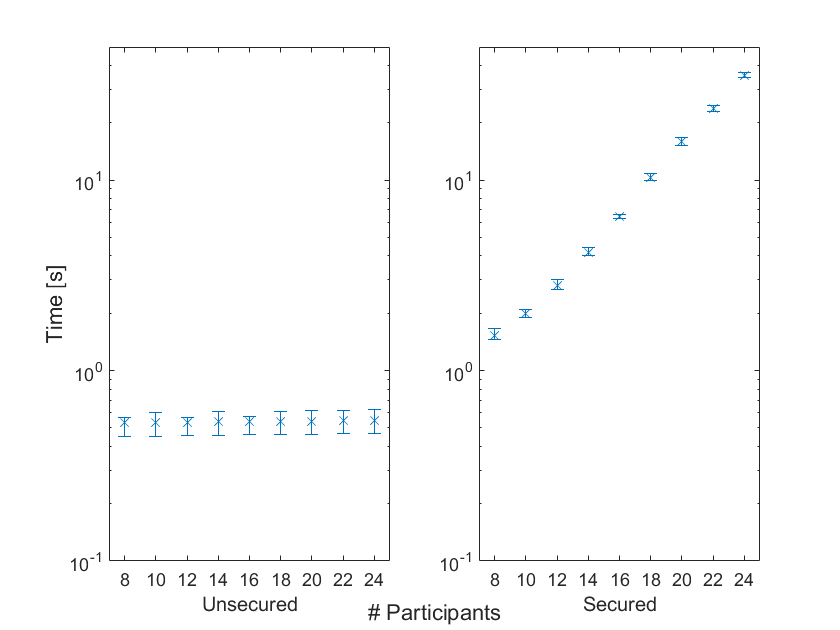}
\caption{Comparisson of minimum, median and maximum runtime for various numbers of participants. Note that the vertical axis is in log-scale.}
\label{fig:plotnodes}
\end{figure}

The observed results are visualised in \Cref{fig:plotnodes}.
It should be noted that the visualisation uses a logarithmically scaled y-axis.
While the runtimes of the secured version significantly increase with the number of participants, the runtimes of the unsecured version remains nearly constant at 0.5s per protocol instance.

For large numbers of participants, we would expect the unsecured version to scale similarly to the secured version.
Though, for the numbers at hand, the bandwidth and computation complexity are not dominant over the transmission latency.
From these results we can derive that with at most 5\% utilisation of the hardware, we can sustain four parallel messages every 10s or \(\frac{\SI{2}{\kibi\byte}}{\SI{10}{\second}}=\SI{204.8}{\byte\per\second}.\)
With full utilisation we reach \(\frac{\SI{2}{\kibi\byte}}{\SI{0.5}{\second}}=\SI{4}{\kibi\byte\per\second}.\)
This speed is sufficient for most text-based applications.

\subsection{Number of Senders and Message Size}

In a second experiment, we looked at the scalability of the protocol with different numbers of senders within a protocol run.
Again, the slot number has been fixed in order to circumvent collisions, which occur if two or more senders randomly select the same slot.
We used a fixed number of $20$ participants and granted each instance four compute threads.
The number of senders $s$ was has been increased from 1 to 20 in increments of 1, which produced effective message sizes of $s\times \SI{512}{\byte}.$
This experiment can be executed through \texttt{messages.sh}.

\begin{figure}[htbp]
\centering
\includegraphics[width=\columnwidth]{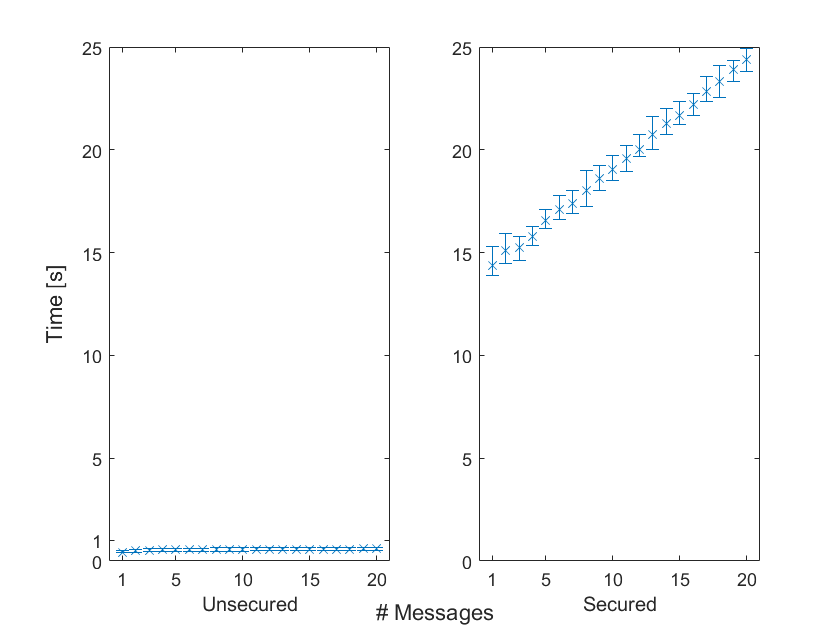}
\caption{Comparison of minimum, median and maximum runtime for various numbers of messages. Overall message length is equivalent to 512 times the number of messages}
\label{fig:plotmessages}
\end{figure}

The results are visualised \Cref{fig:plotmessages}.
Again, it can be observed that the performance of the unsecured variant barely budges under the increasing load. On the other hand, the runtimes of the secured variant increase linear with the number of participants. This is because the number of commitments that have to be generated and validated increases linear with the effective message size, if the number of participants is fixed.
In the optimised unsecured variant, the increase of the effective message size does not increase the runtime of the system noticeably.

We can combine these observations with our results from the previous experiment and improve our previous bound to \(\frac{\SI{10}{\kibi\byte}}{\SI{0.5}{\second}} = \SI{20}{\kibi\byte\per\second}.\)
As the experimental results did not consider larger messages, this provides no upper bound on the transmission efficiency.
For \(n\) participants, the available bandwidth \(b\) of a node will provide an upper limit of \(\frac{1}{2n}b\) to the transmission efficiency of the system, as the message has to be transmitted \(2n\) times for a successful transmission.

\subsection{Performance Optimisation}
\begin{figure}
\centering
\includegraphics[width=0.8\columnwidth]{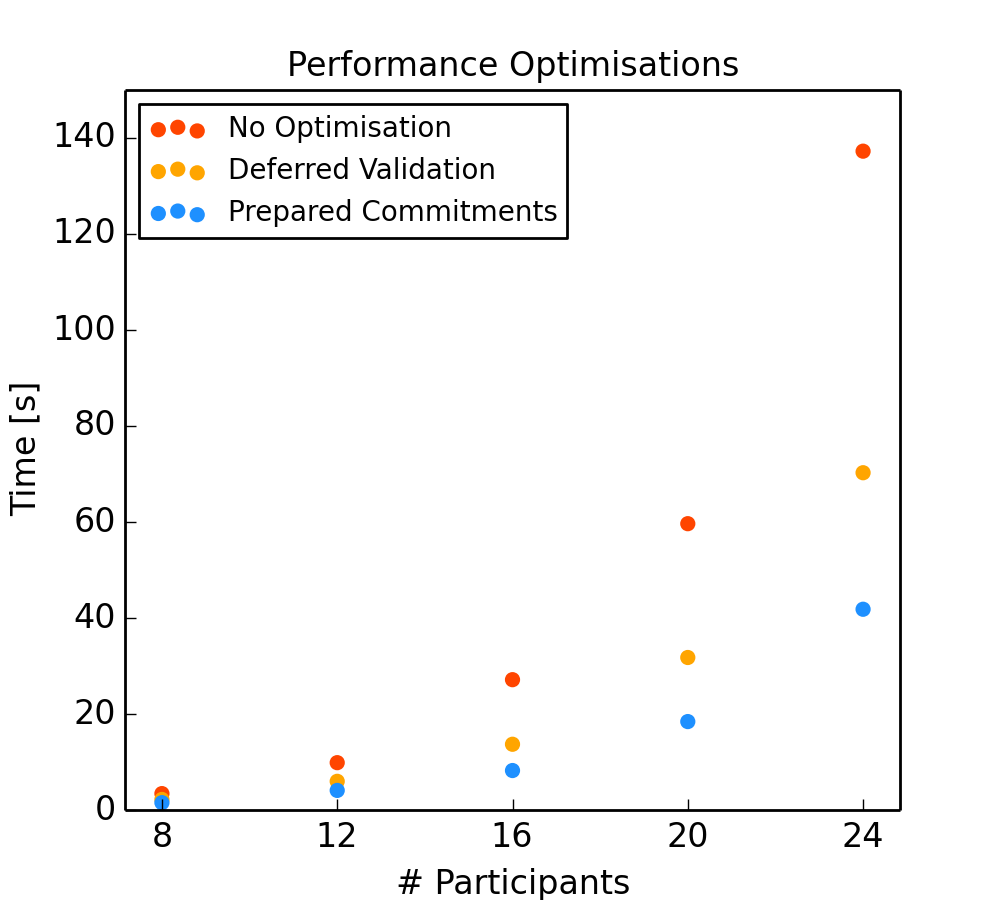}
\caption{Performance impact of the implemented optimisation strategies. Median runtime of several protocol runs in seconds.}
\label{fig:performanceOptimization}
\end{figure}

The performance impact of the previously described optimisation strategies has also been evaluated. 
The experiments were executed with $512$B messages, $k \in \{8, 12, 16, 20, 24\}$ participants and $\frac{k}{2}$ senders.
The results are visualized in \Cref{fig:performanceOptimization}. 
It can be observed that the deferred commitment validation significantly improves the performance, because approximatelly one half of the computationally demanding cryptographic operations can be omitted in normal operation. 
With the additional use of prepared commitments, the performance can be further improved. 
However, the impact of this additional optimisation is smaller, since it only applies to the initial round, because the commitment generation in the final round requires the seeds, distributed in the preceding initial round.

\subsection{Comparison}
\begin{figure*}[t]
	\center
	\includegraphics[width=0.98\textwidth]{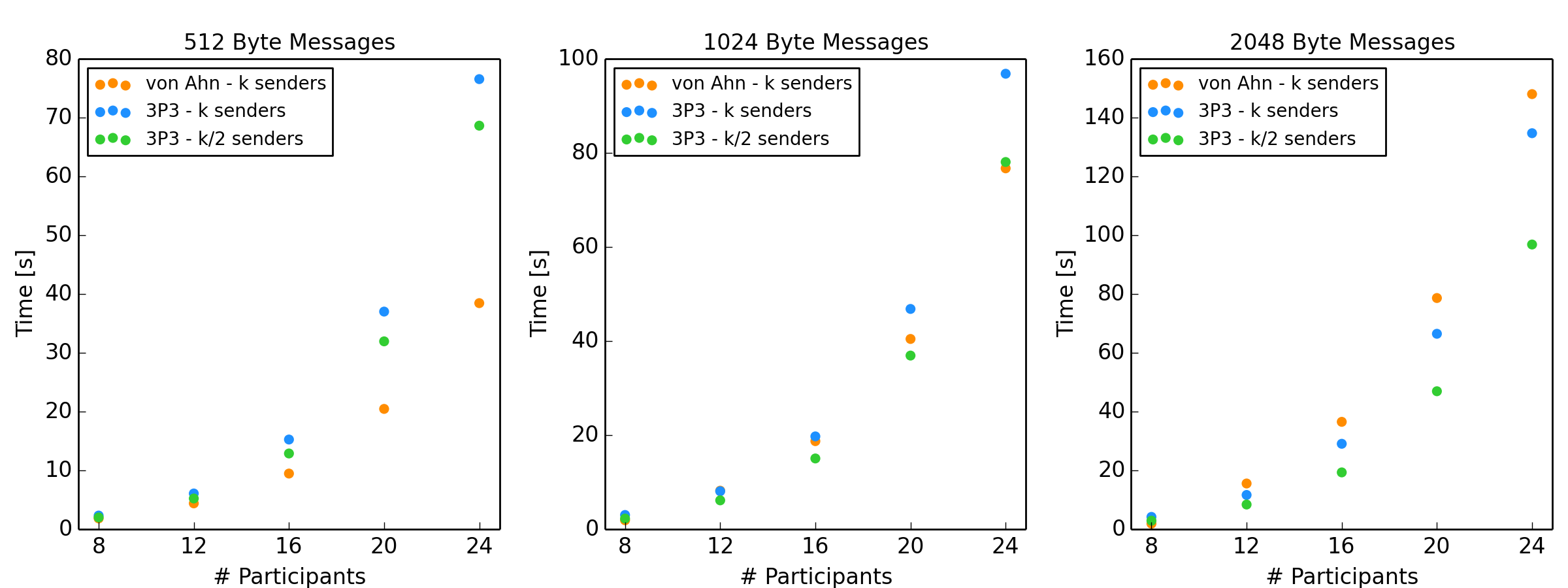}
	\caption{Comparison of our protocol and the von Ahn et al. protocol with three different message sizes. Median runtime of several protocol runs in seconds.}
	\label{fig:vonAhnComparison}
\end{figure*}

In this section, we compare our experimental results to our preliminary results~\cite{moedinger2020trustcom} and discuss the applicability of other proposals.

Our preliminary evaluation~\cite{moedinger2020trustcom} simulated a broadcast using a three phase protocol, where phase one is realized with the k-anonymous message transmission protocol, while phase two and three are used to disseminate the message to the remaining protocol.
The simulation used group sizes between 8 and 20 participants and a varying latency between 20 and 200ms with an expected value of 80ms.
The full protocol showed a 99.9th percentile performance of around 1 second, where phase 2 and 3 are expected to take up to half a second, depending on the network size.
The simulation results are consistent with our experimental results.

An interesting performance comparison for the introduced protocol can be found in Dissent.
While Dissent~\cite{Corrigan2010dissent} fulfils a different purpose than our protocol, it provides similar levels of privacy.
Performance-wise, the core mechanism of dissent requires multiple seconds, up to minutes, for a message round.
The updated version~\cite{wolinsky2012dissentnumbers} shows transmission times from 0.5 to 10 seconds for their message exchange process.
These results use realistic latency distributions through a Planetlab setup, which provides similar latency to our estimated 100ms~\cite{planetlab2014}. 

The proposal by Wang et al.~\cite{wang2007} provides lower latency, 50-100ms, in a 3-hop setup with only 4 to 6 group participants.
Assuming a latency comparable to our setup, the communication latency should increase to at least 330ms, based on the results presented.
Due to the protocol structure, a much higher latency is to be expected.

Finally, we compare the performance of our protocol to the protocol proposed by von Ahn et al. \cite{vonahn2003kanonmessages}. 
Since there is no source code available, accompanying their publication, we implemented our own version that only marginally deviates from their specification. 
Instead of utilizing Pedersen Commitments over large prime fields, we utilize Elliptic Curve Pedersen Commitments similar to our own protocol. 
We executed the experiments with a varying number of participants and three different message sizes. 
\Cref{fig:vonAhnComparison} visualizes the performance of the non-optimised version of our protocol with $k$ and $\frac{k}{2}$ senders and the performance of our implementation of the von Ahn et al. protocol.
While our protocol is slower for $512$B messages, it approaches the performance of the von Ahn et al. protocol with $1024$B messages, and noticeably surpasses its performance with $2048$B messages. 
Additionally, it can be observed that while our protocol scales with the number of senders, the protocol proposed by von Ahn et al. can create significant overhead if the number of parallel senders is low, because its runtime is only affected by the number of participants and the message size used.

\section{Conclusion}
\label{sec:conc}

In this paper, we provided an extension of the protocol by von Ahn et al. to arbitrary length messages.
This allows its application for a sender-k-anonymous broadcast protocol for arbitrary messages.
Further, we optimised the protocol to run faster in its secure mode as well as in a less secure mode for common use-cases, without compromising the privacy of the participants.

We also provided a proof of concept implementation of the resulting protocol.
The optimized secured version requires around $35\si{\second}\pm 2.5\si{\second}$ for the interquartile protocol instances for our largest test.
This test included 24 participants in the network with 4 parallel senders with a total message size of \SI{2}{\kibi\byte} using 4 threads on a participants machine.
The worst outliers took up to \(2.8\times\) the time of the median instances in this setting.

Experiments using this implementation show transmission times of \SI{0.5}{\second} per round based on a \SI{100}{\milli\second} latency for the optimised unsecured round.
The experiments further show little impact of increasing message size or participant numbers in this mode for a reasonable home internet connection, i.e., \SI{50}{\kilo\bit\per\second}.
While it is not an exhaustive analysis, the performance of the secured version should be kept in mind as a fallback mechanism.

The results shown by the secured and unsecured variants of the protocol provide enough throughput for most applications.
The fully secured version is applicable for highly security-relevant applications such as blockchain transactions.
The version only using the secured version as a fallback mechanism can easily be used for many less critical text-based applications.

\bibliographystyle{IEEEtran}
\bibliography{bibliography}

\end{document}